\documentclass[fleqn,twoside]{article}
\usepackage{amsmath}
\usepackage[headings]{espcrc2}

\readRCS
$Id: espcrc2.tex,v 1.2 2004/02/24 11:22:11 spepping Exp $
\usepackage{graphicx}
\usepackage[figuresright]{rotating}

\input babarsym_ams

\def\btn {\ensuremath {\B^+ \to \tau^+ \nu_{\tau}}\xspace}
\def\btodlnux {\ensuremath{\Bub \to \Dz \ell^{-} \bar{\nu}_{\ell} X}}
\def\Vub {\ensuremath{V_{ub}}}

\def\onlumi    {\ensuremath { 288  \invfb\ }}
\def\offlumi   {\ensuremath { 27.5 \invfb\  }}
\def\eextra {\ensuremath{E_{\mathrm{extra}}}}
\def\nBB   {\ensuremath {320 \times 10^{6}} }

\def\nulb  {\ensuremath{\bar{\nu}_{\ell}}}

\def\nutb  {\ensuremath{\bar{\nu}_{\tau}}}

\def\tautoenunu {\ensuremath {\tau^+ \to e^+ \nu_e \nutb}}
\def\tautoe {\ensuremath {\tau^+ \to e^+ \nu_e \nutb}}
\def\enunu {\ensuremath {e^+ \nu_e \nutb}}
\def\tautomununu {\ensuremath {\tau^+ \to \mu^+ \nu_{\mu} \nutb}}
\def\tautomu {\ensuremath {\tau^+ \to \mu^+ \nu_{\mu} \nutb}}
\def\mununu {\ensuremath {\mu^+ \nu_{\mu} \nutb}}
\def\tautopinu {\ensuremath {\tau^+ \to \pi^+ \nutb}}
\def\tautopi {\ensuremath {\tau^+ \to \pi^+ \nutb}}
\def\pinu {\ensuremath {\pi^+ \nutb}} 
\def\tautopipiznu {\ensuremath {\tau^+ \to \pi^+ \pi^{0} \nutb}}

\def\tautorho {\ensuremath {\tau^+ \to \pi^+ \pi^{0} \nutb}}
\def\pipiznu {\ensuremath {\pi^+ \pi^{0} \nutb}}

\title{The Search for $\btn$ at \babar}

\author{Luke A. Corwin (The Ohio State University) for the \babar Collaboration \\
	Stanford Linear Accelerator Center, 2575 Sand Hill Road, Menlo Park, CA 94025
	}
       
\runtitle{$\btn$ at \babar}
\runauthor{L. A. Corwin}

\begin{document}

\begin{abstract}
 We present a search for the decay \btn\ using 288 \invfb of data 
collected at the $\Y4S$ resonance with the \babar\ detector 
at the SLAC PEP-II $B$-Factory. A sample of events with one reconstructed 
semileptonic $B$ decay (\btodlnux) is selected, and
in the recoil a search for $\btn$ signal is performed. The $\tau$ is
identified in the following channels: $\tautoenunu$, $\tautomununu$,
$\tautopinu$ and $\tautopipiznu$. 
We measure a branching fraction of 
$\mathcal{B}(\btn)=(0.88^{+0.68}_{-0.67}(\mbox{stat.}) \pm 0.11 (\mbox{syst.})) \times 10^{-4}$
and extract an upper limit on the branching fraction, at the
90\% confidence level, of $\mathcal{B}(\btn) < 1.8 \times 10^{-4}$.
We calculate the product of the $B$ meson decay constant and $|\Vub|$ to be
$f_{B}\cdot|\Vub| = (7.0^{+2.3}_{-3.6}(\mbox{stat.})^{+0.4}_{-0.5}(\mbox{syst.}))\times10^{-4}$~GeV.
\vspace{1pc}
\end{abstract}

\maketitle

\section{Introduction}

\subsection{Standard Model}
\label{sec:SM_Pred}

We here present a search for the decay $\btn$ at the \babar experiment using the semileptonic tag method \cite{BaBar2006}.  In the Standard Model (SM), purely leptonic decays of the charged $B$ meson proceed via quark annihilation into a $W^{+}$ boson.  The SM branching fraction is given by
\begin{eqnarray}
\BR(B^{+} \rightarrow {\ell}^{+} \nu)&=&
\frac{G_{F}^{2}m_{B}m_{\ell}^{2}}{8\pi} \label{eq:SM_BF}
\\ &\times &\left( 1-\frac{m_{\ell}^{2}}{m_{B}^{2}} \right)^{2} f_{B}^{2} \mid V_{ub} \mid^{2} \tau_{B}, \nonumber
\end{eqnarray}
where $G_F$ is the Fermi constant,
$m_{B}$ and $m_{\ell}$ are the $B$ meson and lepton masses, and $\tau_{B}$ is the B meson lifetime. The branching fraction 
has a simple dependence on $V_{ub}$, the quark mixing matrix element, and 
$f_{B}$, the meson decay constant which describes the overlap of the
quark wave functions within the meson.  Therefore, measuring the branching fraction could provide a clean experimental measurement of the SM value for $f_{B}$.  Note that throughout this document, 
the natural units $\hbar=c=1$ are assumed.
%

Neither this nor any other purely leptonic decay of the charged $B$ meson has been observed.  The most stringent currently published limit is $\mathcal{B}(\btn) < 2.6 \times 10^{-4}$ at the 90\% confidence level \cite{PDG2006}.
%


The SM prediction of \BR($\btn$) can be made in several different ways; we discuss two here.  First, we substitute the experimentally determined values (when possible) and theoretically calculated values (when necessary) for all variables into Equation \ref{eq:SM_BF}.  All values, except $V_{ub}$ and $f_{B}$, are taken from \cite{PDG2006}.  We use $|V_{ub}| = (4.39 \pm 0.33) \times 10^{-4}$ \cite{HFAG2006} and a lattice QCD calculation of $f_{B} =  0.216 \pm 0.022\ \gev$ \cite{Gray2005}.  The result is
\begin{equation}
\BR(\btn) = (1.59 \pm 0.40)\times 10^{-4}.
\label{eq:BF_QCD}
\end{equation}

The second prediction is given by the UT Fitter group:
%
%
\begin{equation}
\label{eq:BF_UT}
\BR(\btn) = (1.41 \pm 0.33)\times 10^{-4}\ \cite{UT2006}.
\end{equation}
These two values agree within uncertainty, and we see that current experimental limits are approaching the SM predictions.

\subsection{Potential New Physics}

In the Type II Two-Higgs Doublet Model (2HDM), purely leptonic charged $B$ decays can proceed via a process identical to the SM process, except that the $W$ is replaced by a charged Higgs boson.  The relationship between the total and SM branching fractions is given by 
\begin{equation}
\frac{\BR(B^{+} \rightarrow {\ell}^{+} \nu)}{\BR(B^{+} \rightarrow {\ell}^{+} \nu)_{SM}}= 
	 \left( 1-tan^{2}\beta \frac{m_{B^{+}}^{2}}{m_{H^{+}}^{2}} \right)^{2},
\end{equation}
where SM denotes the Standard Model branching fraction, $\tan \beta$ is the ratio of the vacuum expectation values of the two doublets, and $m_{H^{+}}$ is the mass of the charged Higgs boson 
\cite{Ho1993}.  

\section{Analysis Methods}
\label{sec:babar}
The data used in this analysis were collected with the \babar\ detector
at the \pep2\ storage ring. 
The sample corresponds to an integrated
luminosity of \onlumi at the \FourS\ resonance (on-resonance) 
and \offlumi \xspace taken $40\mev$ below $B\bar{B}$ threshold 
(off-resonance). The on-resonance sample consists of
$\nBB$  $\FourS$ decays (\BB\ pairs). The collider is operated with asymmetric
beam energies, producing a boost of $\beta\gamma \approx 0.56$ 
of the \FourS\ along the collision axis.

The \babar\ detector is optimized for asymmetric--energy collisions at a
center-of-mass (CM) energy corresponding to the \FourS\ resonance.
The detector is described in detail in Ref.~\cite{NIM2002}. 
The components used in this analysis are the tracking system
composed of a five-layer silicon vertex detector and a 40-layer drift chamber (DCH),
the Cherenkov detector for charged $\pi$--$K$ discrimination, the CsI calorimeter
(EMC) for photon and electron identification, and the 18-layer flux return (IFR) located 
outside of the 1.5T solenoidal coil and instrumented with resistive plate chambers for muon
and neutral hadron identification. For the most recent 51 \invfb of data, a portion of the muon 
system has been replaced with limited streamer tubes~\cite{NIM2006}.
We separate the treatment of the data to account for varying accelerator and detector conditions.
``Runs~1--3'' corresponds to the first 111.9\invfb, ``Run~4'' the following 99.7\invfb , and ``Run~5'' the subsequent 76.8\invfb.

\subsection{Semileptonic Tag}
\label{sec:SemileptonicTag}

In the semileptonic tagging method, one of the two charged $B$ daughters of the \FourS\ (hereafter referred to as the $B^{-}$) is reconstructed in a semileptonic decay mode $\btodlnux$, where $\ell$ is $e$ or $\mu$.  $X$ can be nothing or a transition particle from the decay of a higher mass charm state, which we do not attempt to reconstruct.  Our studies showed that the efficiency gained using this method was worth the signal purity lost by not reconstructing the higher mass state.  

We reconstruct the $\Dz$ candidates in four decay modes:  
$K^{-}\pi^{+}$, $K^{-}\pi^{+}\pi^{-}\pi^{+}$, $K^{-}\pi^{+}\pi^{0}$, and
$K_{s}^{0}\pi^{+}\pi^{-} (K_{s}^{0} \rightarrow \pi^{+}\pi^{-})$. All tag $B$ reconstruction is performed assuming the only unreconstructed particle is the $\nulb$.  We refer to the tracks and neutrals used to reconstruct the tag $B$ as the ``tag side'' of the event. We require the net event charge to be zero and that all tag side tracks meet at a common vertex.

\subsubsection{Double Tag Sample}
\begin{table}[htb]
\caption{Tag Efficiency and Systematic error derived from the Double Tag sample.}
\label{tab:DT}
\renewcommand{\arraystretch}{1.2} 
\begin{tabular}{@{}lll}
\hline
Run & $\frac{\epsilon_{data}}{\epsilon_{MC}}$ & Systematic\\& & Error \\
\hline
	Run 1-3 & $1.05 \pm 0.02$ & 1.9\% \\
	Run 4 & $1.00 \pm 0.03$ & 3.0\% \\
	Run 5 & $0.97 \pm 0.03$ & 3.1\%	\\
\hline
\end{tabular}\\[2pt]
\end{table}
The most important control sample consists of ``double-tagged'' events, for which
both of the $B$ mesons are reconstructed in tagging modes, $\btodlnux$ vs. $\B^+ \to \bar{D}^{0} \ell^+ \nu_{\ell} X$.  This sample has a very similar topology to our signal, and we used it to validate  our Monte Carlo (MC) simulation and correct our tag efficiency.  

For Data and MC, we assume that the number of reconstructed double-tagged events ($N_{2}$) is given by $N_{2} = \epsilon^{2} N$, where $N$ is the total number of $B$ pairs in the data sample and $\epsilon$ is the tag efficiency.  

We calculated $\frac{\epsilon_{data}}{\epsilon_{MC}}$ for Runs 1-3, 4, and 5.  The results are shown in Table \ref{tab:DT}.  The ratio is our correction to the tag efficiency, and the fractional error on the ratio is used as the tag side systematic error.

To validate our MC, we compare double-tagged data events with double-tagged MC events.  The MC samples are weighted so that the number of MC events generated matches the number of data events in the relevant sample. We find excellent agreement in both yield and shape between MC and data.  



\subsubsection{Background Rejection}

For all $\Dz$ decay modes except $K^{-}\pi^{+}\pi^{0}$, the mass of the reconstructed $\Dz$ is required to be within 20 \mev\ of the nominal mass 
\cite{PDG2006}.
In the $K^{-}\pi^{+}\pi^{0}$ decay mode, 
the mass is required to be within 35 \mev\ of the nominal mass
\cite{PDG2006}.  The momentum of the tag lepton in the CM frame is required to be greater than 0.8 \gev. Neutral $B$ mesons are excluded from the tag side by rejecting events with a charged pion that could be combined with a $D^{0}$ to form a $D^{*+}$ candidate.

Assuming that the massless neutrino is the only missing particle, we 
calculate the cosine of the angle between the $\Dz\ell$ candidate
and the $B$ meson,
\begin{equation}
\cos\theta_{B-D^{0}\ell} = \frac{2 E_{B} E_{D^{0}\ell} - m_{B}^{2} - m_{D^{0}
\ell}^{2}}{2|\vec{p}_{B}||\vec{p}_{D^{0}\ell}|}.
\end{equation}
Here ($E_{D^{0}\ell}$, $\vec{p}_{D^{0}\ell}$) and
($E_{B}$, $\vec{p}_{B}$) are the 
four-momenta in the CM frame, and $m_{D^{0}\ell}$ and $m_{B}$ 
are the masses of the $D^{0}\ell$ candidate and $B$ meson, respectively. 
$E_{B}$ and the magnitude of $\vec{p}_{B}$ are calculated 
from the beam energy: $E_{B} = E_{\rm{CM}}/2$ and 
$ | \vec{p}_{B} | = \sqrt{E_{B}^{2} - m_{B}^{2} }$, where 
$E_{B}$ is the $B$ meson energy in the CM frame.

Correctly reconstructed candidates
populate the range [$-1,1$], whereas combinatorial backgrounds
can take unphysical values outside this range. 
We retain events in the interval 
$-2.0 < \cos\theta_{B-D^{0}\ell} < 1.1$, where the upper bound takes 
into account the detector resolution and the loosened lower bound
accepts those events where a soft transition particle from a higher mass
charm state is missing.

\subsection{Signal Side}
\label{sec:Signal_Side}

After excluding the tracks and neutrals 
from the tag side, the remainder of the event (referred to as the ``signal side'')
is searched for consistency with one of the 
above $\tau$ decays, where we require the event contain no extra well-reconstructed tracks. 

Each of the four $\tau$ decay modes used in this analysis produces exactly one charged track.  The well-reconstructed track on the signal side is assigned to a signal category based on its identification as an $e$, $\mu$, or $\pi$.  If a signal $\pi$ can be combined with a $\pi^{0}$ to yield an invariant mass consistent with a $\rho^{\pm}$ (see Table \ref{tab:SigSelSummary}), it is assigned to the $\tautopipiznu$ mode. 

The most important discriminating variable on the signal side is $\eextra$, which is the sum of all detected energy not associated with the tag or signal sides of the event.  For properly reconstructed signal events, this variable should have a peak at zero and a hump at higher values representing the $X$ from the tag $B$.  Thus, this analysis looks for excess events at low values of $\eextra$.  

We ``blind'' the 
signal region of $\eextra$ in data until the final yield
extraction is performed. In this analysis, the blinded region corresponds to $\eextra \leq 500 \mev$; the region where $\eextra > 500 \mev$ is referred to as the ``side band''.  In MC, we calculate the ratio of events observed signal region 
to events observed in the side band.  We multiply this ratio by the number of data events observed in the side band to obtain our background estimate.  The branching fraction and upper limit are extracted from the difference between the background estimate and the observed number of events in the signal region.

\subsubsection{Background Rejection}

\begin{table*}[htb]
\caption{The selection criteria for different signal modes using a $\btodlnux$ tag are listed in this table.}
\label{tab:SigSelSummary}
\renewcommand{\arraystretch}{1.2} 
\begin{tabular}{@{}llll}
\hline
$\tau^+ \to e^+ \nu_e \bar{\nu}_{\tau}$ & 
$\tau^+ \to \mu^+ \nu_\mu \bar{\nu}_{\tau}$ &
$\tau^+ \to \pi^+ \bar{\nu}_{\tau}$ &
$\tau^+ \to \pi^+ \pi^{0} \bar{\nu}_{\tau}$ \\
\hline
$4.6 \le M_{miss} \le 6.7$ & $3.2 \le M_{miss} \le 6.1$ & $1.6 \le M_{miss}$ &   $M_{miss} \le 4.6$  \\
$p^{*}_{signal} \le 1.5$     &   --             & $1.6 \le p^{*}_{signal}$  & $1.7 \le p^{*}_{signal}$      \\
 No IFR \KL & No IFR \KL  & No IFR \KL  & No IFR \KL \\
$2.78 < R_{\tau\tau} < 4.0$ & $2.74 < R_{\tau\tau}$ & $2.84 < R_{\tau\tau}$ & $2.94 < R_{\tau\tau}$  \\
$m_{ee} > 0.1 \gevcc$ & & & \\
$N^{extra}_{\piz} \le 2$ & $N^{extra}_{\piz} \le 2$ & $N_{EMC \KL} \le 2$ & -- \\
     --        &   --         &  --      &  $\rho^{\pm}$ selection:                \\
                 &                &      &  0.64 $< M_{\rho^{\pm}}< $ 0.86 $\gev$   \\ 
                 &                &      &  $0.87 < \cos\theta_{\tau-\rho}$      \\ 

$\eextra < 0.31$ $\gev$ & $\eextra < 0.26$ $\gev$ & $\eextra < 0.48$ $\gev$ &  $\eextra < 0.25$ $\gev$ \\
\hline
\end{tabular}\\[2pt]
\end{table*}

To suppress background we make requirements on the missing mass ($M_{\rm{miss}}$), the momentum of the signal candidate in the CM frame ($p^{*}_{signal}$), the invariant mass of an $e^{-}$ and $e^{+}$ if both are present in our signal side ($m_{ee}$), the number of  $K_{L}$s in the EMC ($N_{EMC \KL}$) and IFR, the number of  extra $\pi^{0}$s in the event ($N^{extra}_{\piz}$), and a continuum background suppression variable denoted $R_{\tau\tau}$.  The values of this requirements are shown in Table \ref{tab:SigSelSummary}. We do not cut on $\eextra$; the values shown in the table define our signal region for each mode.
  
Missing mass is defined as
\begin{equation}
M_{\rm{miss}} = \sqrt{ (E_{\FourS}-E_{\rm{vis}})^2 - ( \vec{p}_{\FourS} - \vec{p}_{\rm{vis}} )^2 }.
\end{equation}
Here ($E_{\FourS}$, $\vec{p}_{\FourS}$) is the four-momentum of the $\FourS$,
known from the beam energies. The quantities $E_{\rm{vis}}$ and $\vec{p}_{\rm{vis}}$ are the 
total visible energy and momentum of the event which are calculated by adding the 
energy and momenta, respectively, of all the reconstructed 
charged tracks and photons in the event.

$R_{\tau\tau}$ is defined as
\begin{eqnarray}
R_{\tau\tau} \equiv &\left[  (3.7-|\cos(\theta_{\vec{T}_{D\ell},\rm{signal}})|)^{2} \right. &
	 \nonumber \\ 
	+  &\left. \left( \frac {M^{\rm{min}}_{3}-0.75\gev}{\gev} \right)^{2}  \right]^{\frac{1}{2}}.&
\end{eqnarray}
It is an empirically derived combination of the cosine of the angle between the signal candidate momentum and the tag $B$'s thrust
vector (in the CM frame), which is denoted $\cos(\theta_{\vec{T}_{D\ell},\rm{signal}})$ and the minimum invariant mass constructible 
from any three tracks in an event (regardless of whether they are already used
in a tag or signal candidates), which is denoted $M^{\rm{min}}_{3}$.
%
%
\section{Results}
\subsection{Yields}
\begin{figure}[htb]
\includegraphics[width=0.8\linewidth]{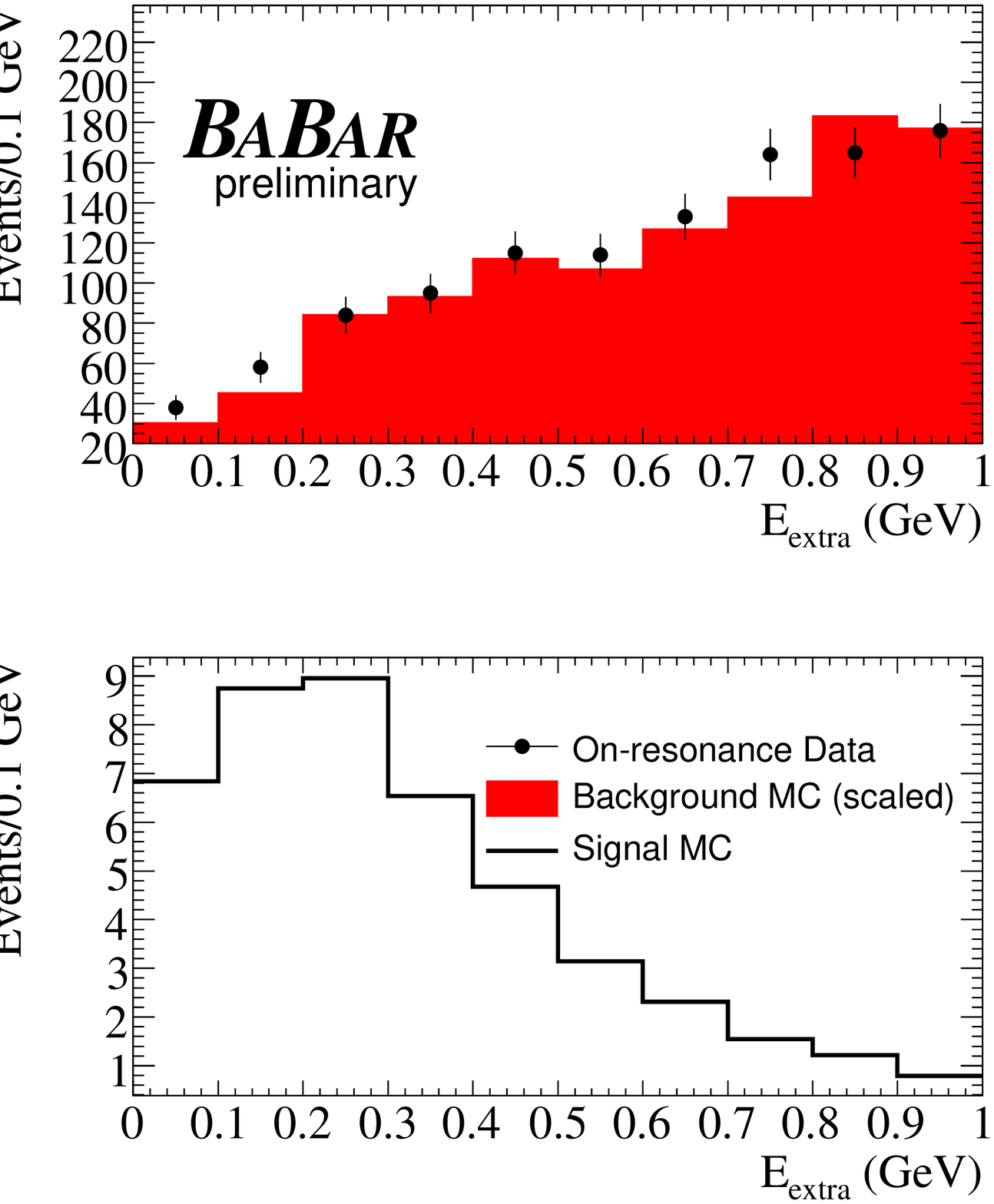}
 \caption{$\eextra$ is plotted after all cuts have been applied with all modes combined. 
MC have been normalized to the on-resonance luminosity. 
In addition, the background MC have been scaled to the data yield in the side band.  Simulated $\btn$ signal MC is plotted (lower) for comparison.}
\label{fig:eextra_allcuts_ALL}
\end{figure}

The unblinded Data and MC are compared in Figure \ref{fig:eextra_allcuts_ALL} and the yields for each mode are shown in Table \ref{tab:unblind-result}.  
\begin{table}[htb]
\caption{The observed number of on-resonance data events in the signal region are shown, together with number of expected background events. The background estimations include all applicable systematic corrections.}
\label{tab:unblind-result}
\renewcommand{\arraystretch}{1.2} 
\begin{tabular}{@{}lll}
\hline
Signal $\tau$        & Expected    & Observed Events  \\ 
Decay                & Background  & in On-resonance \\
Mode                 & Events      & Data  \\ \hline
$\enunu$     & 41.9 $\pm$ 5.2  & 51  \\
$\mununu$    & 35.4 $\pm$ 4.2  & 36  \\
$\pinu$      & 99.1 $\pm$ 9.1  & 109  \\
$\pipiznu$   & 15.3 $\pm$ 3.5  & 17  \\
All modes    & 191.7 $\pm$ 11.8 & 213  \\
\hline
\end{tabular}\\[2pt]
\end{table}
\subsection{Systematic Uncertainty \& Efficiency}

\begin{table*}[htb]
\caption{Contribution to the systematic uncertainty on the signal selection efficiencies in different selection modes.
These uncertainties are added together in quadrature with the uncertainty on the tag $B$ yield, extracted from the
double-tagged control sample, of 1.5\%. The uncertainty on MC statistics is added in quadrature to obtain the total
systematic uncertainty.}
\label{tab:sys_sig}
\renewcommand{\arraystretch}{1.2} 
\begin{tabular}{@{}cccccccc}
\hline
Selection     &  tracking & Particle       & $\KL$  &   $\eextra$  & $\piz$        & Total       &  Correction \\
modes         &   (\%)    & Identification &        &   modeling   & modeling      & Systematic  &  Factor      \\
\hline
$\enunu$      &  0.3     &  2.0            &  3.6    &   3.8      &  --       &     5.8       &  0.982   \\ 
$\mununu$     &  0.3     &  3.0            &  3.6    &   3.8      &  --       &     6.2       &  0.893    \\ 
$\pinu$       &  0.3     &  1.0            &  6.2    &   3.8      &  --       &     7.5       &  0.966    \\ 
$\pipiznu$    &  0.3     &  1.0            &  3.6    &   3.8      &  1.8      &     5.8       &  0.961    \\
\hline
\end{tabular}\\[2pt]
\end{table*}
\babar has an overall systematic uncertainty of 1.1\% on the number of charged $B$ mesons in the data set.  A 1.5\% uncertainty on the tag $B$ yield is estimated using the  double tag sample described in Section \ref{sec:SemileptonicTag}.  The signal side systematic errors are given in Table \ref{tab:sys_sig}.

Overall tag $B$ reconstruction efficiency, which is defined as the fraction of events tagged in a $B\bar{B}$ MC sample with one $B$ decaying to our signal mode and the other decaying generically according to the branching fractions listed in \cite{PDG2006}, is $(6.77 \pm 0.05 \pm 0.10) \times 10^{-3}$.  In this document, in all cases where two uncertainties are quoted on a value, the first uncertainty is statistical and the second is systematic. 
\begin{table}[htb]
\caption{ The signal efficiencies, mode-by-mode, relative to the number of tags. The branching fraction for the 
given $\tau$ decay mode selected is included in the efficiency.}
\label{tab:signal_eff_dlnux}
\renewcommand{\arraystretch}{1.2} 
\begin{tabular}{@{}ll}
\hline
Mode    &    Efficiency (BF Included) \\ 
\hline
$\tautoe$&	0.0414	$\pm$	0.0009 \\
$\tautomu$	&	0.0242	$\pm$	0.0007 \\
$\tautopi$    &	0.0492	$\pm$	0.0010 \\
$\tautorho$	&	0.0124	$\pm$	0.0005 \\
\hline
\end{tabular}\\[2pt]
\end{table}
The selection efficiencies for each signal mode are given in Table \ref{tab:signal_eff_dlnux}, where the efficiency is defined as the ratio of the number of signal events reconstructed (i.e. the rightmost column of Table \ref{tab:unblind-result}) to the number of reconstructed tagged events.  

\subsection{Branching Fraction}
\begin{eqnarray}
	Q \equiv \frac{\mathcal{L}_{s+b}} 
		      {\mathcal{L}_{b}} & CL_{s+b} \equiv P_{s+b}(Q \leq Q_{obs}) \label{eq:CLS1} \\ 
	CL_{s} \equiv \frac{CL_{s+b}}{CL_{b}} & CL_{b} \equiv P_{b}(Q \leq Q_{obs})
	\label{eq:CLS2}
\end{eqnarray}
We use a modified frequentist method, known as the $CL_{s}$ method \cite{Read2002}, for calculating the branching fraction and upper limit.  For this method, we generated a large number of toy MC experiments with different branching fractions in the range from zero to $10 \times 10^{-4}$.  

We define the ``estimator'' $Q$, which is monotonically increasing for increasing signal, and the confidence levels (CL) in Equations \ref{eq:CLS1} and \ref{eq:CLS2}.  $\mathcal{L}$ is the likelihood that a given number of observed events could be produced only by background (denoted with the subscript $b$) or by signal and background (denoted with the subscript $s+b$).  $P_{i}(Q \leq Q_{obs})$ is the fraction of generated toy MC experiments that produced a $Q$ less than or equal to the Q measured in data using a branching fraction that is positive ($i=s+b$) or zero ($i=b$).
\begin{figure}[htb]
\includegraphics[width=35mm]{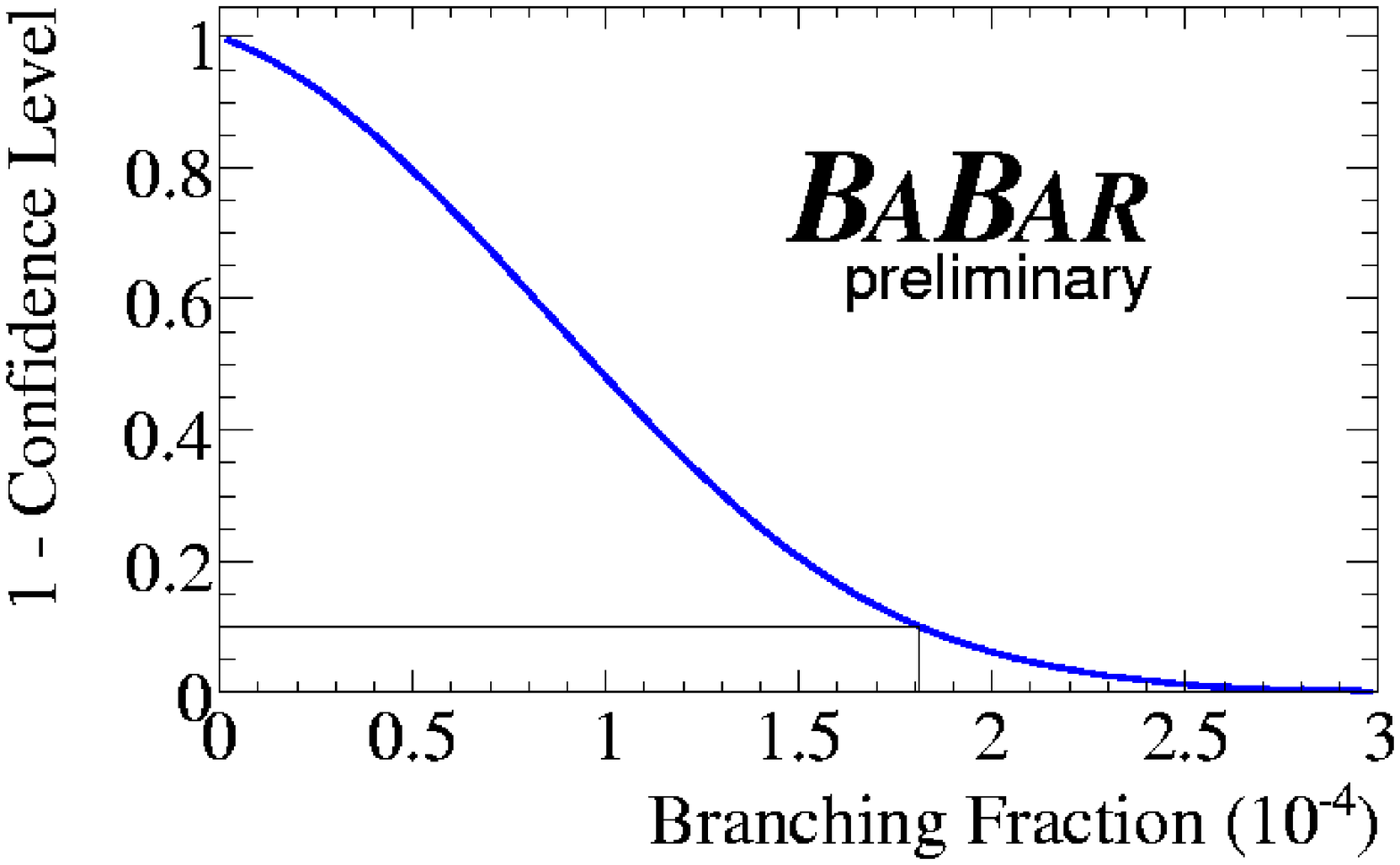}
\includegraphics[width=35mm]{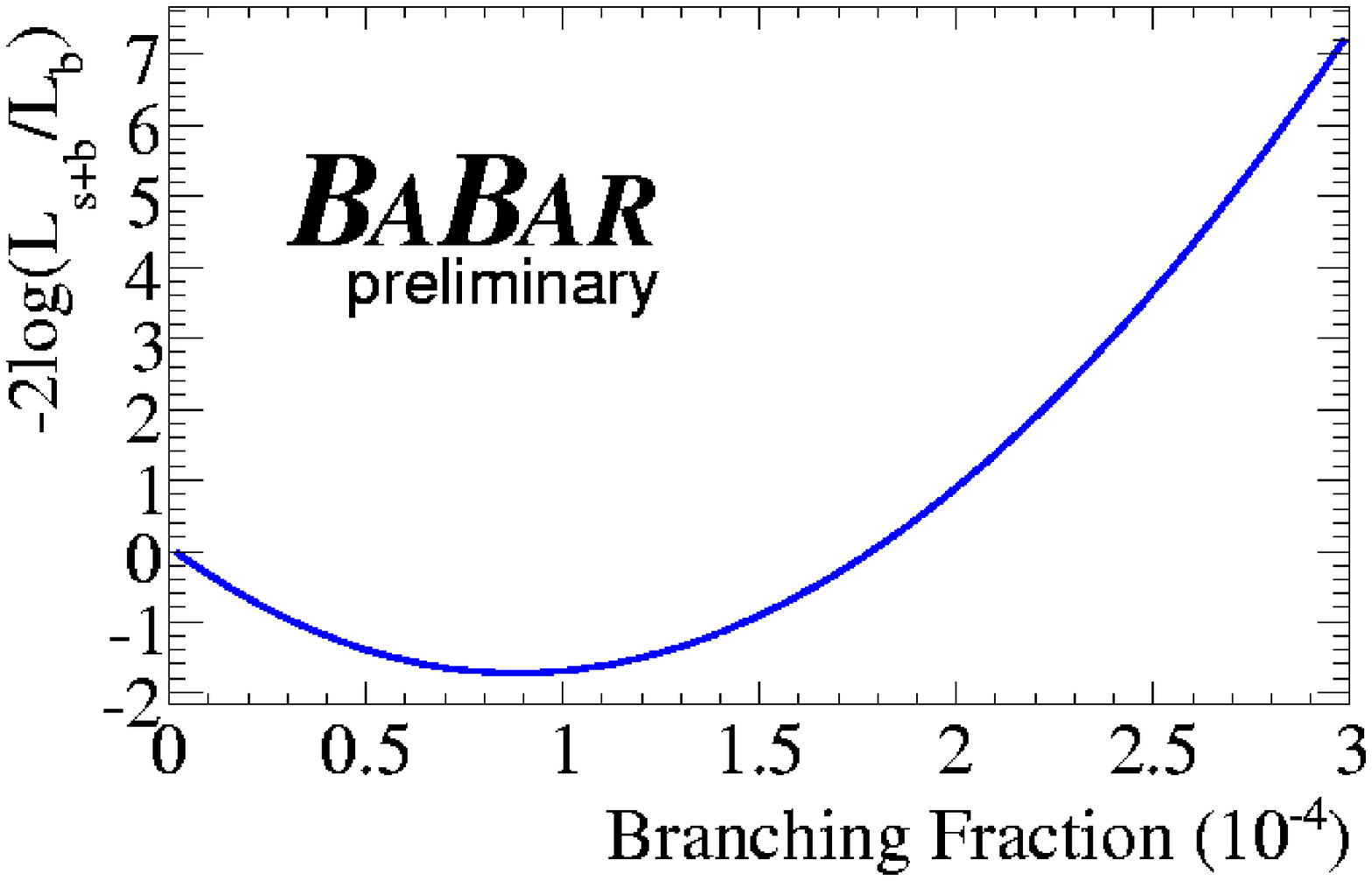}
\caption{The confidence level vs branching fraction is shown (left) to illustrate the extracted upper limit.
The negative log likelihood curve (right) illustrates the central value and it's corresponding uncertainty.}
\label{fig:unblindCL}
\end{figure}
Figure \ref{fig:unblindCL} illustrates our results.  The left-hand plot shows $1-CL_{b}$ as calculated for the number of events expected from different branching fractions.  Our upper limit is the branching fraction for which $CL_{b} = 0.90$.  The right-hand plot shows $-2 \log(Q)$ vs. branching fraction.  Our most likely branching fraction is at the minimum of this curve.  The statistical uncertainty is the distance from the minimum where $-2 \log(Q)$  is 1.0 greater than the minimum.  The resulting branching fraction is only inconsistent with zero by $1.3 \sigma$, so we present it and an upper limit, as well as our result for $f_{B}\cdot|\Vub|$.
\begin{eqnarray}
	\mathcal{B}(\btn)=(0.88^{+0.68}_{-0.67} \pm 0.11 ) \times 10^{-4} \label{eq:BF}\\ 
	\mathcal{B}(\btn) < 1.8 \times 10^{-4} \rm{\ @\ 90\%\ CL}\\ 
	f_{B}\cdot|\Vub| = (7.0^{+2.3\ + 0.4}_{-3.6\ - 0.5})\times10^{-4} \rm{\ GeV} \label{eq:fBVub}
\end{eqnarray}

\subsection{Interpretation \& Conclusion}
Dividing the result shown in Equation \ref{eq:fBVub} by the value of $\Vub$ from \cite{HFAG2006} yields $f_{B} = 0.16^{+0.05}_{-0.08} \gev$, which is consistent with the SM Lattice QCD prediction \cite{Gray2005}.   Our resulting branching fraction is consistent with both predictions shown in Section \ref{sec:SM_Pred} (Equations \ref{eq:BF_QCD} and \ref{eq:BF_UT}).  Therefore, we see no evidence of physics beyond the SM in this analysis.
\begin{figure}[htb]
\includegraphics[width=0.8\linewidth]{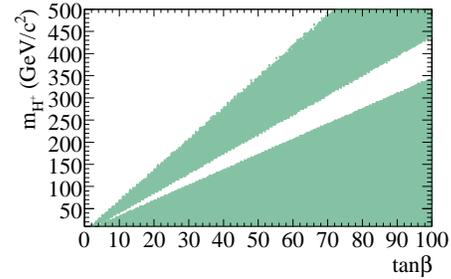}
\caption{The shaded region shows the area excluded by this analysis combined with the search by Belle for the same mode \cite{Belle2006}.  
}
\label{fig:NP_exc}
\end{figure}
This lack of evidence can be translated into bounds on the properties of the charged Higgs boson.  Figure \ref{fig:NP_exc} shows the regions of phase space that can be excluded at the 95\% confidence level in the 2HDM.  Note that  $m_{H^{+}} < 79.3 \gev$ has been excluded by direct searches at LEP \cite{LHWG2001}. 

A search for this mode using hadronic tags is underway at \babar.  We hope to publish a combined result from these two analyses in the near future.

\end{document}